\documentclass{PoS}
\usepackage{epsfig}
\usepackage{cite}
\newcommand{\postscript}[2]{\setlength{\epsfxsize}{#2\hsize}
   \centerline{\epsfbox{#1}}}
\newcommand{\comment}[1]{}

\usepackage[usenames,dvipsnames]{xcolor}
\definecolor{orange}{cmyk}{0,0.5,1,0}
\definecolor{rossoCP3}{cmyk}{0,.88,.77,.40}
\definecolor{graa}{rgb}{0.8,0.8,0.8}
\definecolor{blaa}{rgb}{0.2,0.2,0.6}

\newcommand{\beq}{\begin{equation}}
\newcommand{\eeq}{\end{equation}}
\newcommand{\beqa}{\begin{eqnarray}}
\newcommand{\eeqa}{\end{eqnarray}}
\newcommand{\beqar}{\begin{eqnarray*}}
\newcommand{\eeqar}{\end{eqnarray*}}

\begin{tiny}

\end{tiny} 

\title{The pros and cons of beyond standard model interpretations of ANITA events}

\ShortTitle{Pros and cons of beyond SM interpretations of ANITA events}

\author{L. A. Anchordoqui$^{1,2,3}$, I. Antoniadis$^{4,5}$, V. Barger$^{6}$, F. Cornet$^{7}$, C.~Garc\'{\i}a Canal$^{8}$, M.~Guti\'errez$^{7}$, J. I. Illana$^{7}$, J. G. Learned$^{9}$, D. Marfatia$^{9}$, M.~Masip$^{7}$, S.~Pakvasa$^{9}$, S.~Palomares-Ruiz$^{10}$, J.~F.~Soriano$^{1,2}$, \speaker{T. J. Weiler}$^{11}$\\
$^1$City University of New York, Lehman College, NY, USA; $^2$City University of New York, Graduate Center, NY, USA; $^3$American Museum of Natural History, NY, USA; $^4$LPTHE, Sorbonne Universit\'e, Paris, France; $^5$Albert Einstein Center, Institute for Theoretical Physics University of Bern, Bern, Switzerland; $^6$University of Wisconsin, Madison, WI, USA; $^7$Universidad de
Granada, Granada, Spain; $^8$Universidad Nacional de La Plata, IFLP, CONICET, La Plata, Argentina; $^{9}$University of Hawaii at Manoa, Honolulu, HI, USA; $^{10}$Instituto de F\'{\i}sica Corpuscular (IFIC), CSIC-Universitat de Val\`{e}ncia, Valencia, Spain; $^{11}$Vanderbilt University, Nashville, TN, USA} 

\abstract{The Antarctic Impulsive Transient Antenna (ANITA) experiment
  has observed two air shower events with energy $\sim 500~{\rm PeV}$
  emerging from the Earth with exit angles $\sim 30^\circ$ above the
  horizon. As was immediately noted by the ANITA Collaboration, these
  events (in principle) could originate in the atmospheric decay of an
  upgoing $\tau$-lepton produced through a charged current interaction
  of a $\nu_\tau$ inside the Earth. However, the relatively steep
  arrival angles of these perplexing events create tension with the
  standard model (SM) neutrino-nucleon interaction cross
  section. Deepening the conundrum, the IceCube neutrino telescope and
  the Pierre Auger Observatory with substantially larger exposures to
  cosmic $\nu_\tau$'s in this energy range have not observed any
  events.  This lack of observation implies that the messenger
  particle (MP) giving rise to ANITA events must produce an air shower
  event rate at least a factor of 40 larger than that produced by a
  flux of $\tau$-neutrinos to avoid conflicts with the upper limits
  reported by the IceCube and the Pierre Auger collaborations. In
  addition, the sensitivity of ANITA to MP-induced events must be
  comparable to or larger than those of IceCube and Auger to avoid
  conflict with the non-observation of any signal at these
  facilities. Beyond SM interpretations of ANITA events can be
  classified according to whether the MPs: {\it (i)}~live inside the
  Earth, {\it (ii)}~originate in neutrino-nucleon collisions inside
  the Earth, {\it (iii)}~come from cosmological distances. In this
  communication we investigate the positive and negative facets of
  these three classes of models.}

\FullConference{36th International Cosmic Ray Conference - ICRC2019 -\\
		July 24 - August 1, 2019\\
	 Madison, Wisconsin, USA}

\begin{document}

The ANtarctic Impulsive Transient Antenna (ANITA) has observed two anomalous events, which qualitatively look like air showers initiated by energetic ($\sim 500~{\rm PeV}$) particles that emerge from the ice along trajectories with large elevation angles ($\sim 30^\circ$ above the horizon)~\cite{Gorham:2016zah,Gorham:2018ydl}. As was immediately noted by the ANITA Collaboration, these events may originate in the atmospheric decay of an upgoing $\tau$-lepton produced through a charged current interaction of a $\nu_\tau$ inside the Earth. However, for the angles inferred from ANITA observations, the ice would be well screened from up-going high-energy neutrinos by the underlying layers of Earth, challenging  standard model (SM) explanations~\cite{Romero-Wolf:2018zxt}. A plethora of beyond SM physics models have been proposed to describe ANITA events~\cite{Cherry:2018rxj, Anchordoqui:2018ucj, Huang:2018als, Dudas:2018npp, Connolly:2018ewv, Fox:2018syq, Collins:2018jpg, Chauhan:2018lnq, Anchordoqui:2018ssd, Heurtier:2019git, Hooper:2019ytr, Cline:2019snp, Esteban:2019hcm, Heurtier:2019rkz, Borah:2019ciw}, but systematic effects of data analysis (perhaps induced by transition radiation and/or reflection on analogous subsurface structures) cannot yet be completely discarded~\cite{deVries:2019gzs, Shoemaker:2019xlt}.

The origin for descriptions of ANITA events using transition radiation~\cite{deVries:2019gzs} and reflection without phase inversion on anomalous subsurface structures~\cite{Shoemaker:2019xlt} is a reflected cosmic ray shower. However, as noted in~\cite{Esteban:2019hcm}, if the shower reflection does not take place on a somewhat tilted surface both these ideas appear to be in $2.5\sigma$ tension with the measured polarization angle of ANITA-I event~\cite{Gorham:2016zah}. Moreover, transition radiation predicts that events with large elevations would be anomalous, in slight tension with ANITA data. On the other hand, explanations with reflection without phase inversion on anomalous sub-surface structures require a suppression of the primary surface reflection, which has high dielectric contrast, and would precede any subsurface effect by several ns, enough to get well past the interference regime. In addition, the ANITA effects must be completely broadband, which would tend to erase any quasi-resonant effects such as 2-layer interference. In particular, the ANITA-III event has a clean, undistorted primary pulse, and is accompanied by a clean 30-80~MHz in-phase pulse as well, something very hard to accomplish with anything other than a simple, broadband reflection. The effective bandwidth ratio is of order 700~MHz/40~MHz, more than 15:1. Besides, reflection on subsurface structures predict a fair amount of double events. It is evident that more data are needed to validate or discard these ideas.\footnote{An interesting alternative explored in~\cite{Esteban:2019hcm} is to model the shower reflection with coherent radio waves produced in the ionosphere by physics beyond the SM.}

Beyond SM interpretations of ANITA events can be classified according to whether the {\it messenger particles} (MPs): {\it (i)}~live inside the Earth, {\it (ii)}~originate in neutrino-nucleon collisions inside the Earth, {\it (iii)}~come from cosmological distances. In this paper we investigate the positive and negative facets of these three classes of models.

As speculated in~\cite{Anchordoqui:2018ucj}, the two ANITA events could have similar energies because they result from the two-body decay of a new quasi-stable relic, itself gravitationally trapped inside the Earth. However, the fact that both events emerge at the same angle from the Antarctic ice-cap requires a very {\it atypical} dark matter (DM) density distribution inside the Earth.\footnote{In principle this problem could be avoided in a 2-component DM model endowed with a hidden repulsive interaction which is balanced by the gravitational attraction to favor the required atypical DM distribution inside the Earth.} Moreover, trapping superheavy dark matter (SHDM) relics in the required amount is also a rather non-trivial problem for this type of {\it (i)} model~\cite{Dudas:2018npp, Cline:2019snp, Anchordoqui:2018qom}.

For type {\it (ii)} MP production, an extraterrestrial flux of high energy neutrinos is required. We can classify the source distribution producing these neutrinos into two categories. We can have a diffuse flux from steady-state sources isotropically distributed over the sky, or else a non-isotropic flux from transient sources. The non observation of any similar event by IceCube~\cite{Aartsen:2018vtx} and/or the Pierre Auger Observatory~\cite{Aab:2019auo} makes the isotropic source distribution unlikely. More concretely, in the energy range of interest, the exposure to tau neutrinos collected by ANITA is about a factor of 40 smaller than the one collected by IceCube or Auger~\cite{Romero-Wolf:2018zxt}. On the other hand, if the neutrino flux is generated by extremely high-luminosity EeV $\nu$-transients the IceCube and Auger diffuse bounds can indeed be evaded.  However, any cosmic population of such transients (assumed to be extragalactic) must be isotropic, ongoing, and include sources exhibiting a broad range of fluxes here at Earth, due both to source distance effects and any associated luminosity function. No such cosmic population of high-luminosity, high-frequency ($\gg 2~{\rm month}^{-1}$, all-sky) neutrino transients can be compatible with the limits on neutrino point sources~\cite{Aartsen:2017kru}, for $E_\nu \gtrsim 200~{\rm TeV}$~\cite{Aartsen:2017kru} and $E_\nu \gtrsim 1~{\rm PeV}$~\cite{Aartsen:2018vtx} neutrinos, and rates of neutrino multiplet events~\cite{Aartsen:2017snx} set by IceCube. A way around these constraints is to have a high-multiplicity extreme-energy neutrino interaction, such that the energy of the $\tau$-lepton is one to two orders of magnitude smaller than the neutrino energy. However, for $E_\nu \gtrsim 10^{10.6}~{\rm GeV}$, ANITA sets the most restrictive bounds on the astrophysical neutrino flux~\cite{Gorham:2019guw}, and therefore the IceCube and Auger diffuse bounds could be evaded.

By far the most attractive proposal in this category is the production of a stau or bino in $R$-symmetric SUSY models (in which the gravitino is the lightest supersymmetric particle)~\cite{Fox:2018syq,Collins:2018jpg}. This idea, originally proposed in~\cite{Albuquerque:2003mi}, has been extensively studied in the literature~\cite{Albuquerque:2006am,Ahlers:2006pf,Ahlers:2007js,Ando:2007ds}. The main problem with this hypothesis is that the neutrino-nucleon cross
section for production of SUSY particles is about 4 orders of magnitude smaller than the SM charged-current (CC) interaction, and so one can only accommodate ANITA data if the extraterrestrial flux of neutrinos originates in transient
sources~\cite{Collins:2018jpg}. However, non-perturbative processes may come to the rescue: SUSY sphaleron transitions have a cross section about an order of magnitude larger than that of CC interactions~\cite{Anchordoqui:2018ssd}. An advantage here is that the high-multiplicity final state acts as an amplifier of MPs with respect to $\nu$'s, and in principle one can have the required factor of 40 if all SUSY fermions decay to the next-to-lightest supersymmetric
particle~\cite{Cerdeno:2018dqk}. Besides, all SUSY fermions and neutrinos with energy above the sphaleron barrier would interact again producing more of these transitions, while degrading the energy and increasing the SUSY-$\nu$ ratio. The energy of the observed events by ANITA roughly coincides with that of the sphaleron barrier~\cite{Moreno:1996zm}. It is of interest to investigate the region of the SUSY parameter space that favors the elevation angles observed by ANITA. It is this
that we now turn to study.

\begin{figure}[tpb]
\begin{minipage}[t]{0.31\textwidth}
\postscript{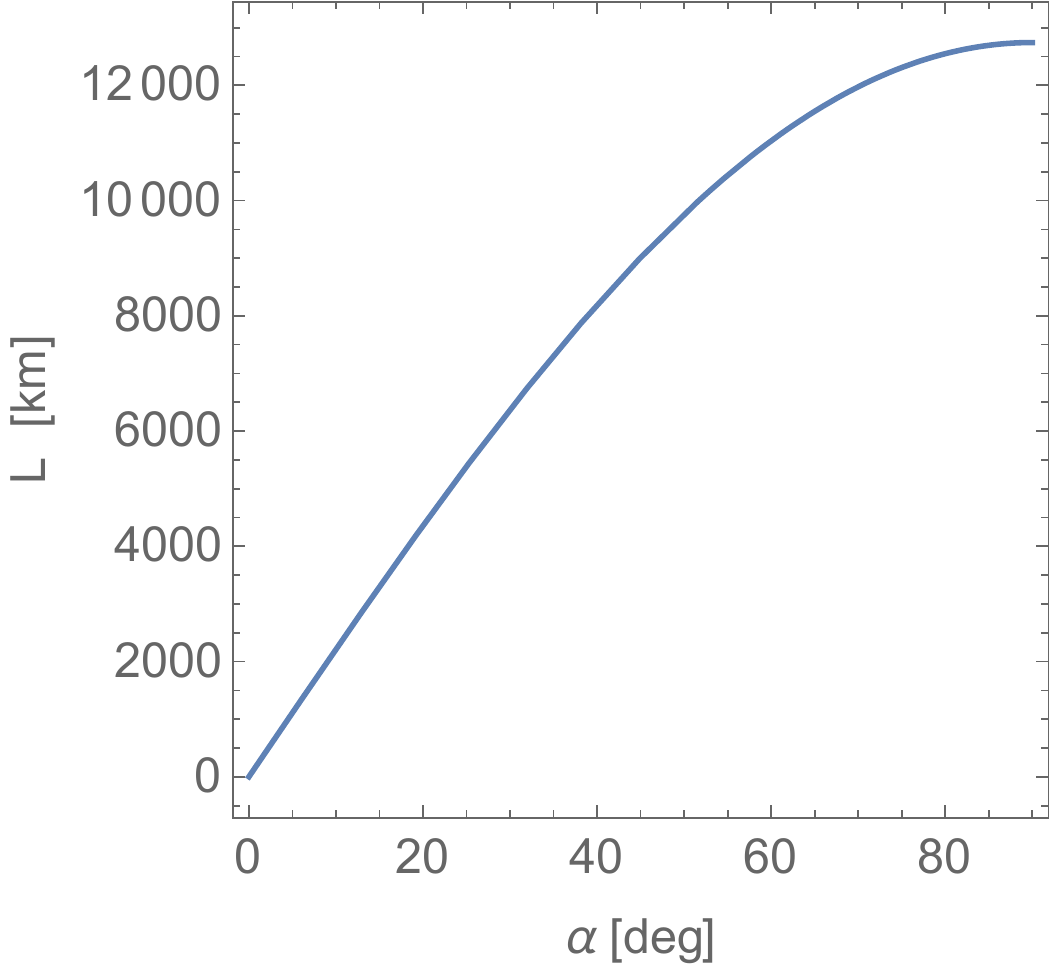}{0.98}
\end{minipage}
\begin{minipage}[t]{0.31\textwidth}
\postscript{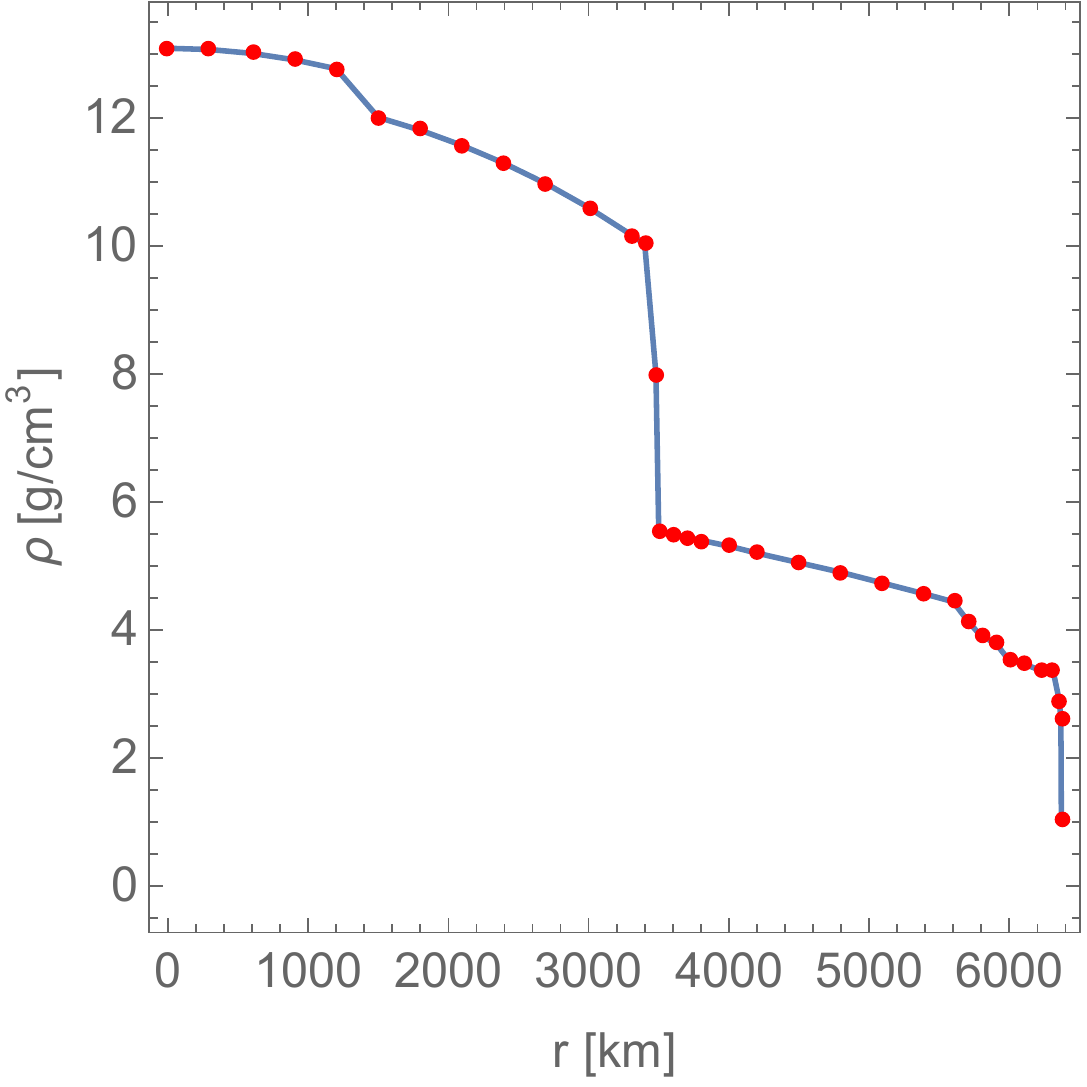}{0.90}
\end{minipage}
\begin{minipage}[t]{0.31\textwidth}
\postscript{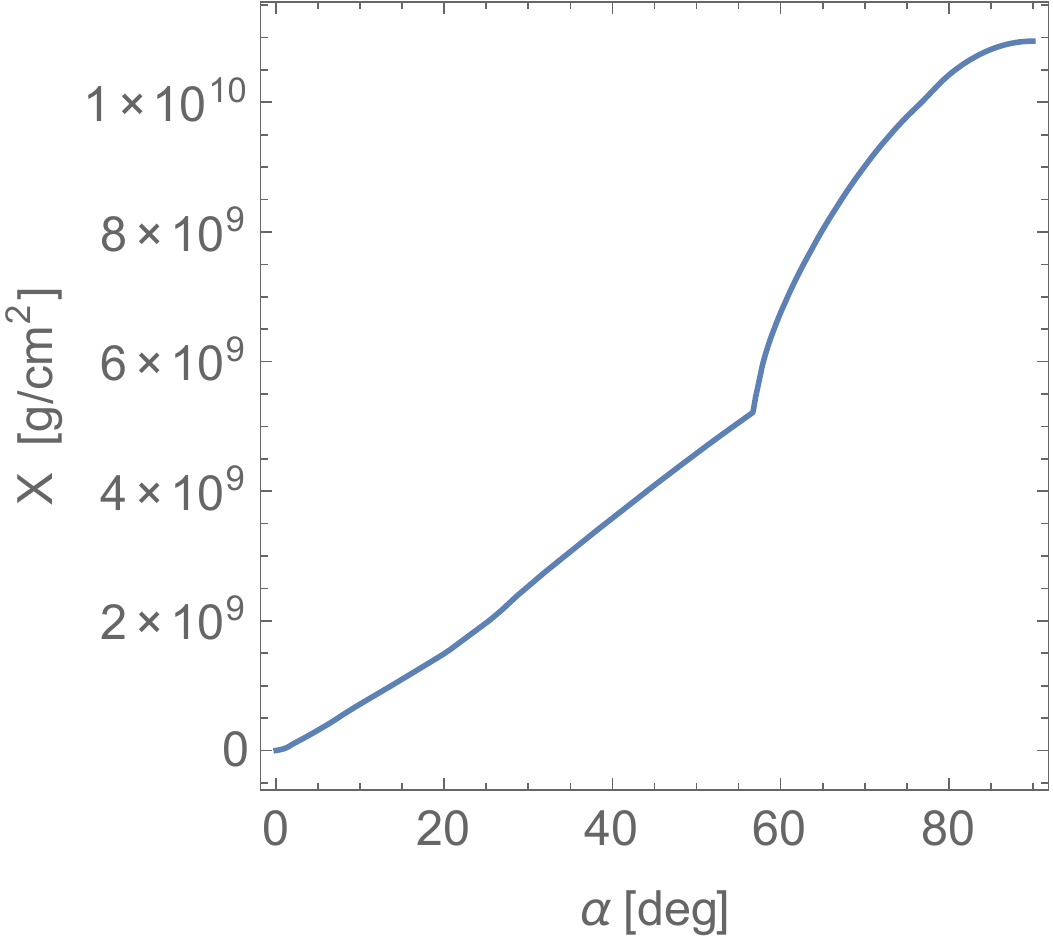}{1.00}
\end{minipage}
\caption{{\it (a)}~Chord length $L$ as a function of the exit angle $\alpha$ . {\it (b)}~Earth's density profile. {\it (c)} Total  column depth $X(\alpha)$.
\label{fig:1}}
\end{figure}

Consider an stau of mass $m$ and energy $E$ produced by a neutrino near the Earth's surface. The incoming direction of the neutrino defines a trajectory of chord $L$. A particle following that trajectory will emerge with an exit angle $\alpha$ respect to the horizontal. The chord length as a function of $\alpha$ is shown in Fig.~\ref{fig:1}.a. Now, we can use the Earth density shown in Fig.~\ref{fig:1}.b to transform the chord length $L(\alpha)$ into the total depth $X(\alpha)$ that the stau faces before emerging. The corresponding total column depth as a function of $\alpha$ is displayed in Fig.~\ref{fig:1}.c.

\begin{figure}[tpb]
\begin{minipage}[t]{0.31\textwidth}
\postscript{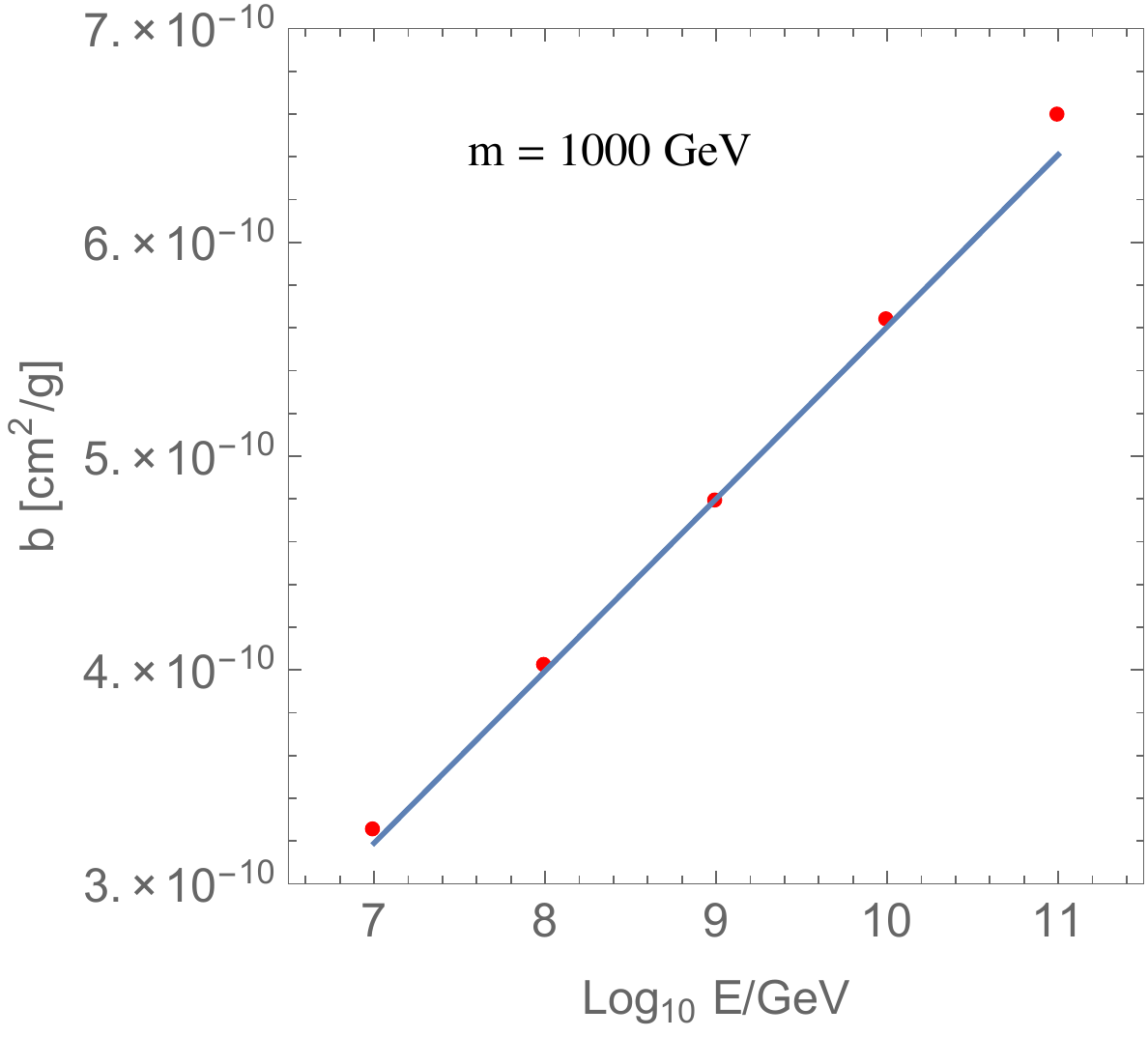}{0.96}
\end{minipage}
\begin{minipage}[t]{0.31\textwidth}
\postscript{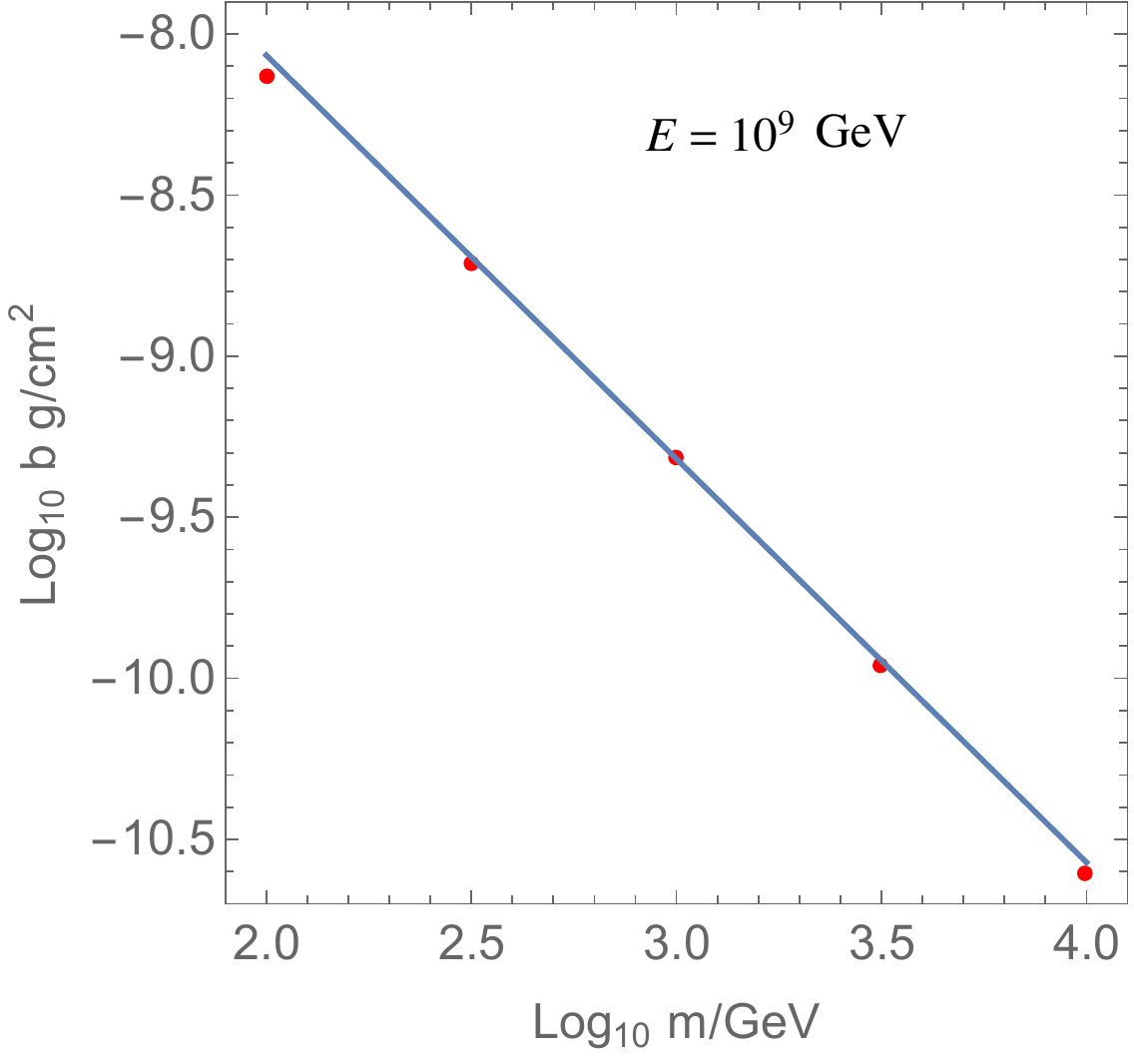}{0.90}
\end{minipage}
\begin{minipage}[t]{0.31\textwidth}
\postscript{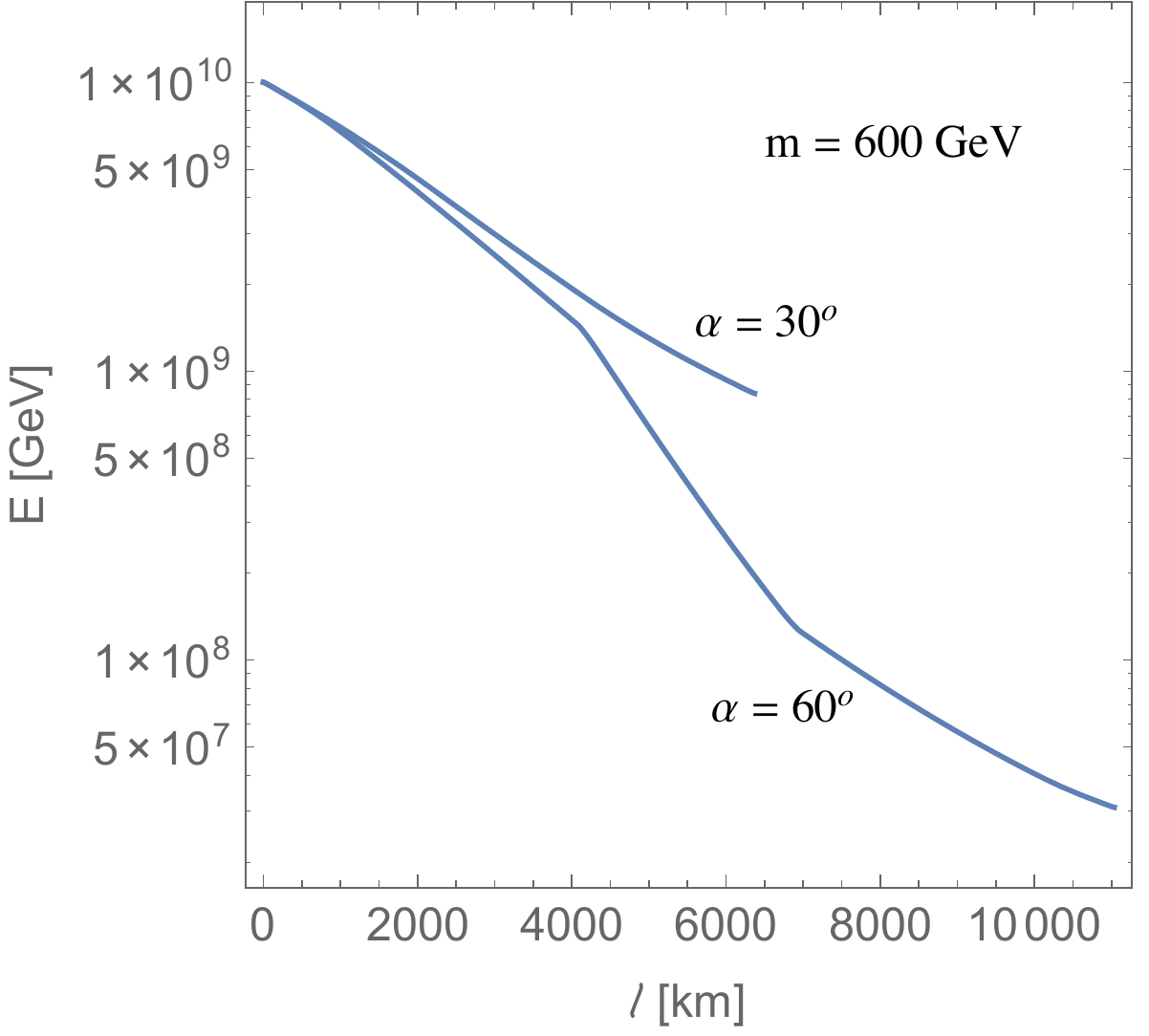}{0.93}
\end{minipage}
\caption{{\it (a).}~Fraction of energy lost by a stau per unit depth as a function of $E$ for $m = 1~{\rm TeV}$. {\it (b).}~Fraction of energy lost by the stau per unit depth as a function of $m$ for $E = 10^9~{\rm GeV}$. {\it (c).}~Energy degradation $E$ as a function of the distance $l$. \label{fig:2}}
\end{figure}

For $10^2 \lesssim m/{\rm GeV} \lesssim 10^4$ and $10^7 \lesssim E/{\rm GeV} \lesssim 10^{11}$, the fraction of energy lost by the stau per unit depth of standard rock can be parameterized as, \beq b(m,E)=4.8 + 10^{-10}\,\left( {10^3\,{\rm GeV}\over m}\right)^{1.25} \left( 1+0.073\, \log {E\over 10^9\,{\rm GeV}}\right);
\label{eq:b}
\eeq see Figs.~\ref{fig:2}.a and \ref{fig:2}.b where we compare the results from numerical integration with the parametrization in (\ref{eq:b}).  The degradation of the stau energy is given by 
\beq
{{\rm d}E \over {\rm d} l} = - \,b(m,E)\; \rho(l)\; E\,,
\label{el}
\eeq
where $l$ goes from 0 to $L(\alpha)$. As an illustration, in Fig.~\ref{fig:2}.c we show the energy degradation for $m=600$~GeV and two inclination angles.

\begin{figure}[tpb]
\begin{minipage}[t]{0.49\textwidth}
\postscript{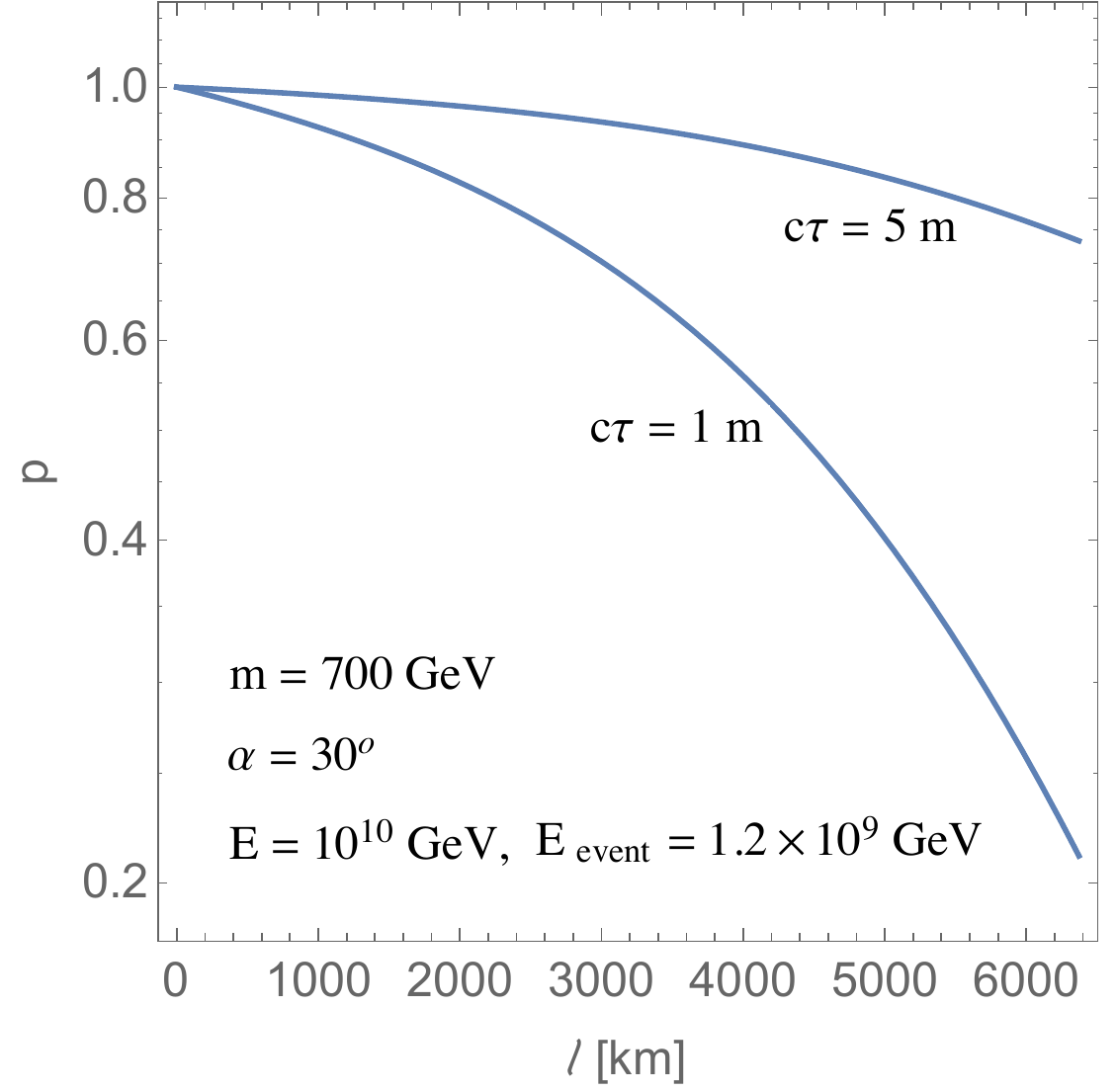}{0.925}
\end{minipage}
\begin{minipage}[t]{0.49\textwidth}
\postscript{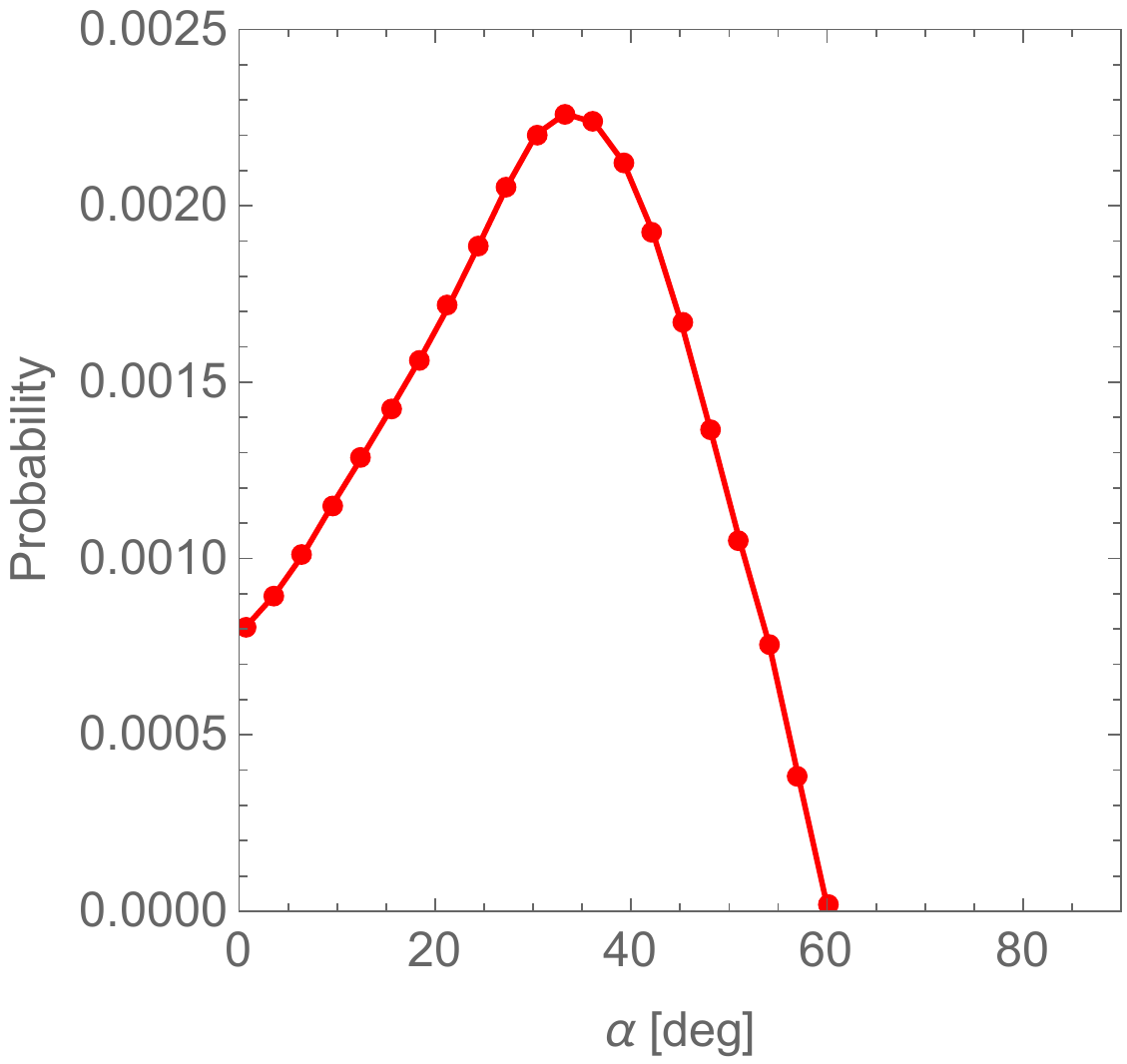}{0.98}
\end{minipage}
\caption{{\it (a)} Stau survival probability as a function of distance $l$. {\it (b)} Detection probability as a function of the exit angle $\alpha$ for $E = 5\times 10^9~{\rm GeV}$, $m=800~{\rm GeV}$, and $c\tau=4~{\rm m}$.\label{fig:3}}
\end{figure}

We now turn to the probability for the stau to emerge. The stau decay length is 
\beq 
\lambda_{\rm dec}(\tau,m,E) = {c \,\tau\,E\over m}\,\sqrt{1-{m^2\over E^2}} ~. 
\eeq
As the stau crosses the Earth, it may decay. The probability $p(l)$ that it survives along its trajectory satisfies 
\beq 
{{\rm d}p \over {\rm d} l} = - {p\over \lambda_{\rm dec}} ~.  
\eeq 
Since we already know $E(l)$ from (\ref{el}), we can integrate this equation numerically and obtain the probability to have the stau at $l=L(\alpha)$; examples are shown in Fig.~\ref{fig:3}.a. Finally, if the stau emerges with energy $E$, the probability that it decays within 20 km (so that ANITA can see it) is just
\beq
p_{\rm event}=1-\exp \left( -{20\,{\rm km}\over \lambda_{\rm dec}} \right) ~.
\eeq
Note that staus with large lifetimes would be able to cross the Earth, but they would have a small probability to decay to form the anomalous ANITA events. And staus with short lifetimes would  decay prematurely while crossing the Earth.

We can now scan to search for optimal parameters. The problem has 4 input and 2 output variables: initial energy of the stau ($E$), mass of the stau ($m$), its lifetime ($\tau$) and inclination angle ($\alpha$) as inputs; probability ($p_{\rm event}$) and energy ($E_{\rm event}$) of the stau event as outputs. Herein, we consider the simplest scenario with all the energy of the emerging stau going into the shower energy of the event. After scanning we find a region of the parameter space that favors the $\alpha \sim 30^\circ$ chord through the Earth. For example, if we fix the input parameters to $E = 5\times 10^9~{\rm GeV}$, $m=800~{\rm GeV}$, and $c\tau=4~{\rm m}$, in agreement with LHC searches~\cite{Aaboud:2018jbr}, the probability has a maximum $p_{\rm event} \sim 0.22\%$ at $\alpha=34^\circ$, with a final shower energy $E_{\rm event}=7\times 10^8~{\rm GeV}$ consistent with ANITA events; see Fig.~\ref{fig:3}.b.

MPs proposed in type {\it (iii)} models are sterile neutrinos~\cite{Cherry:2018rxj, Huang:2018als} or some relativistic particle from the hidden sector produced via SHDM decay in the Galactic halo~\cite{Hooper:2019ytr, Cline:2019snp, Heurtier:2019rkz}. The hybrid model in which SHDM decays into a pair of right-handed sterile neutrinos that mutate into active $\nu_\tau$'s in their passage through the Earth has also been explored~\cite{Heurtier:2019git}.

Regardless of the nature of the source, sterile neutrinos would lose their coherence after propagation for a long distance producing mass eigenstate fluxes $\nu_4$ and $\nu_1$, with fractions of $\cos^2 \vartheta$ and $\sin^2 \vartheta$, respectively, where $\vartheta$ is the active-sterile mixing angle, $\nu_1$ stands for an active neutrino mass eigenstate, and $\nu_4$ represents the heaviest, keV, neutrino. The results of a parameter space scan indicate that to accommodate observation of two events at ANITA the flux of $\nu_4$ should saturate the IceCube limit $d\Phi_{\nu_4}/d\Omega \approx 2 \times 10^{-15} [0.1/\sin \vartheta]^2 ({\rm cm^2 \, s \, sr})^{-1}$, with $m_4 \gtrsim 1~{\rm keV}$ and $\vartheta \lesssim 0.1$~\cite{Huang:2018als}. Note that if SHDM decays were the sources of these sterile neutrinos, the mixing angle could not be vanishingly small, because it would be too difficult to produce the required neutrino flux that saturates the IceCube bound. For parameters required to observe the two anomalous events by ANITA, the predicted event rate at IceCube is a factor of $\sim$6 larger.

Models in which SHDM decays to a highly boosted, light DM particle, which can interact in the Earth to produce $\tau$ leptons require some fine-tuning~\cite{Cline:2019snp}. The level of fine-tuning is reduced if the weakly interacting DM particles scattering elastically with nuclei in the Antarctic ice sheet produce Askaryan emission (similar to Cherenkov in a dense dielectric medium)~\cite{Hooper:2019ytr}. It is important to stress that while the measured waveforms and polarizations angles of ANITA's anomalous events are inconsistent with Askaryan emission from neutrino induced showers, they can be consistent with Askaryan emission from showers produced in the interactions of exotic weakly interacting particles, to which the Earth is transparent. Needless to say, the smoking gun for models of SHDM decaying in the halo of the Milky Way would be a directional signal towards the Galactic Center. None of the observed anomalous ANITA events point to that direction. Moreover, as recently shown in~\cite{Chipman:2019vjm}, isotropic fluxes of active/sterile neutrinos or other exotic particles that couple to the $\tau$ lepton through suppressed weak interaction cross sections cannot be responsible for the anomalous ANITA events.

In summary, we have reviewed a variety of models that can partially accommodate ANITA observations, but a convincing unified explanation of all data is yet to see the light of day. Hence, obviously we need more data. The second generation of the Extreme Universe Space Observatory (EUSO) instrument, to be flown aboard a super-pressure balloon (SPB) is under construction to fly from Wanaka (New Zealand) by 2022~\cite{Adams:2017fjh}. EUSO- SPB2 will look down on the atmosphere with an optical fluorescence detector from the near space altitude of 33~km and will look towards the limb of the Earth to observe the Cherenkov signal of cosmic rays from above the limb and cosmic neutrino showers generated just below the limb. EUSO-SPB2 will provide an important test both of the anomalous ANITA events and the various beyond SM physics models discussed in this paper.\\

We thank Peter Gorham and David Saltzberg for valuable discussions. This work has been supported by NASA (80NSSC18K0464), U.S. NSF (PHY-1620661), DoE (DE-SC-0010504, DE-SC-001198), Institute Lagrange de Paris, Swiss NSF, CNRS PICS, AN- PCyT, MICINN of Spain (FPA2016-78220, FPA2017-84543-P and SEV-2014-0398), Junta de Andalucia (FQM101), and European Union's Horizon 2020 program (Marie Sk\l odowska-Curie No. 690575 and 674896).

\end{document}